\documentclass[aps.pra,onecolumn,showpacs,preprintnumbers,amsmath,amssymb]{revtex4}

\usepackage[utf8]{inputenc}
\usepackage[english]{babel}

\usepackage{setspace}
\usepackage{booktabs}
\usepackage{multirow}
\usepackage{amsmath,amssymb}
\usepackage{qcircuit}
\usepackage{caption}
\usepackage{lipsum}
\usepackage{natbib}
\usepackage{graphicx}
\usepackage{subcaption} 

\usepackage{physics}
\usepackage[figurename=Fig., justification=RaggedRight]{caption}
\usepackage{amsmath}
\usepackage{xcolor}
\usepackage{hyperref}

\begin{document} 

\title{Depth analysis of variational quantum algorithms for heat equation}

\author{N. M. Guseynov$^{1,2}$, A. A. Zhukov$^{1}$, W. V. Pogosov$^{1,2,3}$,A.V. Lebedev$^{1,2}$}
\affiliation{$^1$Dukhov Research Institute of Automatics (VNIIA), Moscow, 127030, Russia}
\affiliation{$^2$Moscow Institute of Physics and Technology (MIPT), Dolgoprudny, 141700, Russia}
\affiliation{$^3$Institute for Theoretical and Applied Electrodynamics, Russian Academy of Sciences, Moscow, 125412, Russia}

\begin{abstract}
Variational quantum algorithms are a promising tool for solving partial differential equations. The standard approach for its numerical solution are finite difference schemes, which can be reduced to the linear algebra problem. We consider three approaches to solve the heat equation on a quantum computer. Using the direct variational method we minimize the expectation value of a Hamiltonian with its ground state being the solution of the problem under study. Typically, an exponential number of Pauli products in the Hamiltonian decomposition does not allow for the quantum speed up to be achieved. The Hadamard test based approach solves this problem, however, the performed simulations do not evidently prove that the ansatz circuit has a polynomial depth with respect to the number of qubits. The ansatz tree approach exploits an explicit form of the matrix what makes it possible to achieve an advantage over classical algorithms. In our numerical simulations with  up to $n=11$ qubits, this method reveals the exponential speed up.
\end{abstract}

\maketitle

\section{INTRODUCTION}

Quantum computing is a promising technology based on the principles of quantum mechanics \cite{FEYNMAN}. The main motivation is outperform the state-of-the-art  classical algorithms and achieve the so called quantum supremacy \cite{q_c_nowadays,Q_supremacy1}. Well-known  examples are the quantum search algorithm \cite{grover,q_search} and Shor's factorization algorithm \cite{Shor,math_apply}, which are both superior to the classical ones. These algorithms however require large-depth quantum circuits composed of highly accurate quantum gates. Both these requirements are problematic in the present era of noisy intermediate-scale quantum (NISQ) computers \cite{preskill2018quantum}. Error correction codes \cite{errors,Noh2020FaulttolerantBQ,Darmawan2021PracticalQE,Egan2020FaultTolerantOO} and error mitigation techniques \cite{temme2017error,endo2018practical,endo2019mitigating,sun2021mitigating,zhukov2022quantum} could potentially overcome these problems, however, the state-of-the-art quantum devices lack enough number of qubits to work in the fault-tolerant regime.
Quantum computers can also be useful in various linear algebra problems. A remarkable example is the Harrow, Hassidim, and Lloyd (HHL) algorithm for solving  systems of linear equations \cite{hhl,montanaro2016quantum,wang2020quantum,lee2019hybrid}. Because the classical algorithms generally have the polynomial complexity in the matrix size $N$,  HHL algorithm provides the exponential speed-up in the case of sparse matrices. However, the use of the HHL algorithm for heat equation and similar problems suffers from the exponential decay of the success probability due to the sequential application of the algorithm for each time step.

An alternative approach to the solution of linear equations on quantum computers is variational algorithms. The essence of such algorithms is a combination of the classical minimization for some objective functions whose values are found by means of the quantum computer. In particular, the variational algorithms have demonstrated their effectiveness in the presence of noise \cite{cerezo2021variational,gonzalez2021pricing,alghassi2022variational,wang2021can,fontanela2021quantum,fontana2021evaluating,kubo2021variational,lubasch2020variational,yang2021variational,liu2021variational,bravo2019variational,radha2021quantum}. However, there are limitations to the effectiveness of the objective function minimization, including noise-induced barren plateaus \cite{wang2021noise,cerezo2021cost}, as well as the suppression of cost gradient magnitudes \cite{holmes2022connecting}.

In this paper, we study variational algorithms for solving systems of linear equations obtained by a finite difference scheme of the heat equation at a single time step. We focus on a possibility to achieve quantum superiority using the quantum variational approach. We analyze the optimal choice of the time evolution partition and have found that the time grid parameter controls the computational speed-up in comparison with the classical algorithms. We also found that there are two main problems for the efficient implementation of variational algorithms for solving a linear system: (i) the measurement of the loss function and (ii) the construction of the variational ansatz. The direct variational approach is based on the minimization of the expectation value of some auxiliary Hamiltonian \cite{xu2021variational,sato2021variational}. This approach demonstrates the fundamental possibility of solving systems of linear equations on a quantum computer, but suffers from both mentioned problems. The Hadamard test based approach \cite{xu2021variational} solves the first problem of the loss function measurement by using a specific measurement basis. However, we have not achieved, in this case, an efficient preparation of the ansatz circuit with the help of the universal entanglers. The ansatz tree approach \cite{huang2021near} is based on the given form of a linear system for constructing the ansatz. This approach combines both the efficient measurement of the loss function by exploiting the Hadamard test and the efficient construction of the ansatz by using the special hierarchical optimization technique. This allows us to achieve a superiority over the classical algorithm. The performance of discussed algorithms is demonstrated by the numerical simulation with the matrix size up to $2048\times2048$ ($2^{11} \times 2^{11}$).

The article is organized as follows. In Section II we introduce the heat
equation and its discrete finite difference scheme. The direct variational  algorithm is introduced and discussed in Section III. 
Section IV introduces and describes the Hadamard test based approach and  presents its numerical implementation on a simulator. In Section V we investigate the implementation of the ansatz tree approach. The analysis of the computational complexity of the algorithm is presented. Section VI discusses the optimal choice of the time grid parameter for the heat equation. Section VII concludes the paper with final remarks. 



\section{Heat equation}

Let us consider the heat equation with constant coefficients, a given initial time constraint, and periodic spatial boundary condition,
\begin{eqnarray}
a^2 \Delta U(\vec{r},t)-\frac{\partial U(\vec{r},t)}{\partial t}=f(\vec{r},t);\qquad
U(\vec{r},0)=\chi(\vec{r}); \qquad U(\vec{r},t)=U(\vec{r}+\vec{R},t),
\label{Thermal_conductivity_equation}
\end{eqnarray}
where $f(\vec{r}, t)$ and $\chi(\vec{r})$ are the heat sources and initial distribution function, respectively. In this paper we address the numerical solution of the Eq.(\ref{Thermal_conductivity_equation}). We first analyze the one-dimensional case and then discuss  generalization to the multidimensional situation. The one-dimensional equation reads as
\begin{eqnarray}
a^2\frac{\partial^2 U(z,t)}{\partial z^2}-\frac{\partial U(z,t)}{\partial t}=f(z,t),\qquad
U(z,0)=\chi(z),\qquad U(z,t)=U(z+Z,t).
\label{1_dimensional_thermal_conductivity_equation}
\end{eqnarray}

For the numerical solution of such equations, a grid of values of the arguments of $U$ is introduced. The values of the function $U(z,t)$ on the grid are used for the construction of the finite difference scheme \cite{zienkiewicz2005finite,ozicsik2017finite} based on the equation
\begin{eqnarray}
a^2\frac{U^{\tau+1}_{i+1}-2U_i^{\tau+1}+U_{i-1}^{\tau+1}}{(\delta z)^2}-\frac{U^{\tau+1}_i-U^\tau_i}{\delta t}=f^\tau_i;\qquad
U^0_i=\chi_i, \qquad U^\tau_0=U^\tau_{N_z}
\label{Thermal_conductivity_equation_grid_initial_decomposition}
\end{eqnarray}
which can be rewritten as
\begin{eqnarray}
(-2-c)U^{\tau+1}_i+U^{\tau+1}_{i+1}+U_{i-1}^{\tau+1}=b_i^\tau,\qquad
U_i^0=\chi_i,\qquad U_0^\tau=U_{N_z}^\tau,
\label{equation_grid}
\end{eqnarray}
where
\begin{eqnarray}
c=\frac{(\delta z)^2}{a^2\delta t}, \qquad b^\tau_i=(f_i^\tau+\frac{U_i^\tau}{\delta t})\frac{(\delta z)^2}{a^2},
\label{c_definition}
\end{eqnarray}
where $\delta z $ and $\delta t$ are the spatial and temporal partitions of the grid, $N_z$ and $N_t$ are the numbers of space and time partitions, respectively. Index $i$ corresponds to the coordinate grid, and $\tau$ corresponds to the time grid.
We use an implicit difference scheme that yields the stability of the solution for arbitrary parameters of the equation and the grid size \cite{higham2002accuracy}. Thus, Eq. (\ref{1_dimensional_thermal_conductivity_equation}) turns into Eq. (\ref{equation_grid}), where $b^\tau_i$ contains information about previous time layer. The single time step evolution from $t=\tau$ to $t=\tau+1$ is carried out by solving the linear system (\ref{equation_grid}) relative to the spatial coordinates at the time instance $\tau+1$, where the information from the previous time layer $\tau$ is considered to be known through the function $b_i^\tau$. The solution of the system (\ref{equation_grid}) is equivalent to the solution of a system of linear equations
\begin{eqnarray}
Ax=b,
\label{Ax=b}
\end{eqnarray}
where
\begin{eqnarray}
A(c)=\left(\begin{array}{cccccc}
-2-c & 1 & 0 & \dotsm & 0 & 1 \\
1 & -2-c & 1 &\dotsm& 0 & 0\\
0 & 1 & -2-c &\dotsm&  0 & 0\\
\rotatebox[origin=c]{270}{\dots}&&\rotatebox[origin=c]{-45}{\dots}&\rotatebox[origin=c]{-45}{\dots}&&\rotatebox[origin=c]{270}{\dots}\\ 
0 & 0 &\dotsm& -2-c & 1 &0\\
0 & 0 &\dotsm& 1 & -2-c &1\\
1 & 0 &\dotsm& 0 & 1 &-2-c\\
\end{array}
\right),
\label{A_definition}
\end{eqnarray}
\begin{eqnarray}
x=\left(\begin{array}{c}
U_0^{\tau+1}  \\
U_1^{\tau+1} \\
\rotatebox[origin=c]{270}{\dots}\\
U_{N-1}^{\tau+1}\\
\end{array}
\right),\qquad b=\left(\begin{array}{c}
b_0^{\tau}  \\
b_1^{\tau} \\
\rotatebox[origin=c]{270}{\dots}\\
b_{N-1}^{\tau}\\
\end{array}
\right).
\label{Axb}
\end{eqnarray}
The matrix $A$ depends on  the spatial boundary conditions. In this paper we consider the periodic boundary spatial conditions, where the matrix $A$ has the form (\ref{A_definition}).

\subsection{Analysis of the condition number}
The loss of the accuracy due to arithmetic operations in the classical numerical solution of a system of linear equations is governed by the condition number \cite{belsley2005regression}
\begin{eqnarray}
\kappa(A)=\frac{\lambda_{max}(A)}{\lambda_{min}(A)},
\label{condition_number_definition}
\end{eqnarray}
where $\lambda_{min}(A)$, $\lambda_{max}(A)$ are the absolute minimum and maximum eigenvalues. This parameter defines the sensitivity of the vector $b$  to a change in the vector $x$. For example, $\kappa(A)=10^k$ means the loss of $k$ digits in accuracy due to arithmetic operations. The condition numbers for a matrix and its inverse coincide. 

At $\kappa(A)\rightarrow+\infty$ the linear system is unstable with respect to small perturbations in the initial conditions and the solution is not correct at all. Let us consider the condition number for the matrix $A$, given by Eq.(\ref{A_definition}). In this case $\lambda_{max}(A)=c+4$, $\lambda_{min}(A)=c$ and the condition number is given by
\begin{eqnarray}
\kappa(A)=\frac{c+4}{c};\qquad\lim_{c\rightarrow 0}\kappa(A)=+\infty.
\label{condition_number_for_matrix_A}
\end{eqnarray}
Hereafter we assume that the number of coordinate partitions $N_z$ is fixed, and therefore the grid parameter $c$ is determined by the number of partitions of the time interval $N_\tau$. With $c\to 0$ the condition number diverges, and the system (\ref{Ax=b}) becomes unstable at small $c$.

\subsection{Analysis of the precision of the classical numerical solution for the heat equation}

In this subsection we discuss the choice of the optimal grid parameter $c$ for the classical numerical solution of the heat equation given in the paper. The optimal $c$ is chosen through the minimization of the total error, which is comprised of two separate errors: (i) the error of the derivatives approximation by the finite differences, (ii) the arithmetic error due to a finite accuracy of the arithmetic operations.

The arithmetic error is determined by the condition number. At each next time step of the algorithm $\hat{U}_i^\tau \to \hat{U}_i^{\tau+1}$ the solution $\hat{U}_i^{\tau+1}$ inherits the accumulated error $\epsilon_\tau$ of the solution from the previous step $U_i^\tau$ amplified by the conditional number, see  Eq. (\ref{c_definition}), and gets the additional arithmetic error $\tilde{\epsilon}$ resulting from the finite accuracy of the floating-point operations.  Thus, the error for $N_{\tau}\propto c$ time steps can be expressed as
\begin{eqnarray}
\begin{gathered}
\epsilon_0=0;\\
\epsilon_1=\widetilde{\epsilon},\\
\epsilon_2=\sqrt{\widetilde{\epsilon}^2+(c\kappa(A)\epsilon_1)^2},\\
\epsilon_{N_\tau}=\sqrt{\widetilde{\epsilon}^2+\sum_{k=1}^{N_\tau}(\kappa(A)c)^{2k}\epsilon^2_{k-1}}=\widetilde{\epsilon} \sqrt{\frac{(c\kappa(A))^{2N_\tau}-1}{(c\kappa(A))^{2}-1}} \sim \widetilde{\epsilon} (c+4)^{N_\tau-1}.
\end{gathered}
\label{error_conditional_number}
\end{eqnarray}
Equation (\ref{error_conditional_number}) shows that the arithmetic error grows as $\mathcal{O}(c^{N_{\tau}})$ with the number $N_{\tau}\propto c$ of time partitions, while the time derivative approximation error behaves as $\mathcal{O}(1/c)$ according to Ref. \cite{morton2005numerical}. From the behavior of both errors, it can be concluded that there is an optimal choice of the time partition number $N_\tau$ that minimizes the total error.

\section{Direct variational method}

In this section we use the variational method \cite{xu2021variational} to solve a system of linear algebraic equations (\ref{Ax=b}) with the matrix $A$ of the size $2^n \times 2^n $, $n>1$, given by Eq. (\ref{A_definition}). The main idea of the method is to construct a Hamiltonian with a ground state corresponding to the linear algebra problem to be solved.

Suppose we need to solve a system of equations (\ref{Ax=b}). A solution can be formally written as
\begin{eqnarray}
\ket{x}=A^{-1}\ket{b}.
\end{eqnarray}
It can be readily shown \cite{xu2021variational} that this solution $\ket{x}$ is the ground state of the Hamiltonian
\begin{eqnarray}
H=A^{+} (I - \ket{b}\bra{b})A,
\label{var_ham}
\end{eqnarray}
and $E=\bra{x}H\ket{x}=0$.

Thus, the idea of the variational method is to construct a parameterized quantum circuit, which is referred later as the ansatz
\begin{eqnarray}
U_\phi=U_\phi(\vec{\theta}):U_\phi(\vec{\theta})\ket{0}=\ket{\phi(\vec{\theta})}.
\label{ansatz0}
\label{parametrized _quantum_circuit}
\end{eqnarray}
The minimal expectation value of the Hamiltonian (\ref{var_ham}) is found by tuning the parameters $\vec{\theta}$ of the quantum state $\ket{\phi(\vec{\theta})}$
\begin{eqnarray}
\vec{\theta}_m=arg\,min \bra{\phi(\vec{\theta})}H\ket{\phi(\vec{\theta})}.
\label{theta=argmin_ham}
\end{eqnarray}
In this case, $\ket{x}=\ket{\phi(\vec{\theta}_m)}$ is the solution. This problem with the matrix size $2^n \times 2^n $ can be solved on a quantum computer using $n$ qubits. For this purpose it is necessary to set the variational function $\ket{\phi(\vec{\theta})}$ and then measure the expectation value of the Hamiltonian. To measure the expectation value, it can be decomposed into Pauli products:
\begin{eqnarray}
H=\sum_i c_i \sigma_1 \sigma_2 \dots \sigma_n,
\label{pauli_string_1}
\end{eqnarray}
where $\sigma$ are the Pauli matrices $I, \sigma_x, \sigma_y, \sigma_z$.

Then the Anzatz quantum circuit is run for each Pauli product and the expectation value $E=\bra{x}H\ket{x}$ is evaluated. In the standard variational procedure the variational parameters are changed using a classical computer and the procedure is repeated until zero energy is reached with a given accuracy. As a result, we obtain a set of parameters $\vec{\theta}_m $ which correspond to the desired solution $\ket{x}$.

Demonstration of the method was performed on a real IBM quantum computer via QISKIT \cite{IBMQ}. The size of the matrix $A$ (Eq.~\ref{A_definition}) was chosen $4 \times 4$, where only two qubits are required. Qubits 0 and 1 of the ibmq\_manila processor are used in this work. Quantum processor characteristics are given in Table 1. 

\begin{figure}[ht]
\centering
\resizebox{\columnwidth}{!}{%
\begin{tabular}{|l|l|l|l|l|l|ll|}
\hline
~ & $T_1$; µs & $T_2$; µs & $U_2$ gate error & $U_2$ gate duration; ns & Resonant frequency; GHZ & \multicolumn{1}{l|}{Readout error} & Id gate error rate \\ \hline
Qubit 0 & 126.98 & 91.92 & 0.000204 & 35.556 & 4.963 & \multicolumn{1}{l|}{0.0221} & 0.0002 \\ \hline
Qubit 1 & 155.43 & 69.83 & 0.000214 & 35.556 & 4.838 & \multicolumn{1}{l|}{0.0235} & 0.0002 \\ \hline
~ & $R_z$ error rate & $S_x$ error rate & X error rate & Reset; ns & Cnot (C:0; T:1) error rate & \multicolumn{2}{l|}{Cnot (C:1; T:0) error rate} \\ \hline
Qubit 0 & 0 & 0.0002 & 0.0002 & 5514,67 & \multirow{2}{*}{0.00567} & \multicolumn{2}{l|}{\multirow{2}{*}{0.00567}} \\ \cline{1-5}
Qubit 1 & 0 & 0.0002 & 0.0002 & 5514.67 & ~ & \multicolumn{2}{l|}{}\\ \hline
\end{tabular}}
\caption*{Table 1. Average characteristics of the ibmq\_manila processor during measurements.}
\end{figure}
For demonstrative purposes, let us consider $c=0$ (Poisson equation), and assume that $b$ is real. In this case $\det A=0$ and the linear system (\ref{Ax=b}) is not determined. The eigenvector corresponding to $\lambda=0$ is homogeneous: $\vec{e}_{\lambda=0}=\{1,1,1,1\}/\sqrt{4}$. Thus undo an additional constraint  $\vec{b}\cdot \vec{e}_{\lambda=0}=0$ corresponding to $\sum b_i=0$ the linear system (\ref{Ax=b}) has infinitely many solutions $\vec{x} = A^{-1} \vec{b} + \alpha \vec{e}_{\lambda=0}$ with an arbitrary real $\alpha$. We construct $\ket{x}$ using an additional constraint 
\begin{eqnarray}
\begin{gathered}
\ket{x}: \sum x_i=0,
\label{b=0}
\end{gathered}
\end{eqnarray}
which fixes $\ket{x}$ and implies that the solution has no homogeneous Fourier component: $(\vec{e}_{\lambda=0}\cdot \vec{x} = 0$. Hence, two parameters are sufficient to determine the variational function $\ket{x}=\ket{\phi(\theta_1,\theta_2)}$. 

\begin{figure}
\begin{center}
\includegraphics[width=0.99\textwidth]{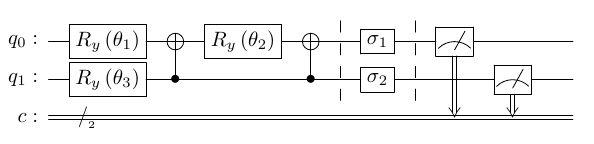}
\caption{\label{circ1}Quantum circuit for the direct variational algorithm for solving the system of linear equations on two qubits. $\sigma_1, \sigma_2$ correspond to the measured Pauli product. Parameters $\theta_1, \theta_2, \theta_3$ specify the variational function $\ket{x}=\ket{\phi(\vec{\theta}_m)}$. }
\end{center}
\end{figure}

The quantum circuit is shown in Fig.~\ref{circ1}. Three parameters $\theta_1, \theta_2, \theta_3$ provide a variational function $\ket{x}$, with only the first two of them being independent, and the third one according to the condition (\ref{b=0}) reads as
\begin{eqnarray}
\begin{gathered}
\theta_3=-2 \arctan \frac{\cos{\theta_+} +\sin{\theta_+}}{\cos{\theta_-} +\sin{\theta_-}},\\
\theta_{\pm}=\frac{1}{2}\left( \theta_1 \pm \theta_2 \right).
\end{gathered}
\end{eqnarray}
The gates $\sigma_1, \sigma_2$ are chosen depending on the Pauli products term~(\ref{pauli_string_1}):
\begin{eqnarray}
\begin{gathered}
\sigma_x \rightarrow H, \sigma_y \rightarrow S^+H,\sigma_z \rightarrow I,
\end{gathered}
\end{eqnarray}
where 
\begin{eqnarray}
H=\frac{1}{\sqrt{2}}\left
(
\begin{array}{cc}
1 & 1 \\
1 &-1 
\end{array}
\right),\quad
S^+=\left
(
\begin{array}{cc}
1 & 0 \\
0 &-i 
\end{array}
\right).
\label{HSgate}
\end{eqnarray}

\begin{figure}
\begin{center}
\includegraphics[width=0.45\textwidth]{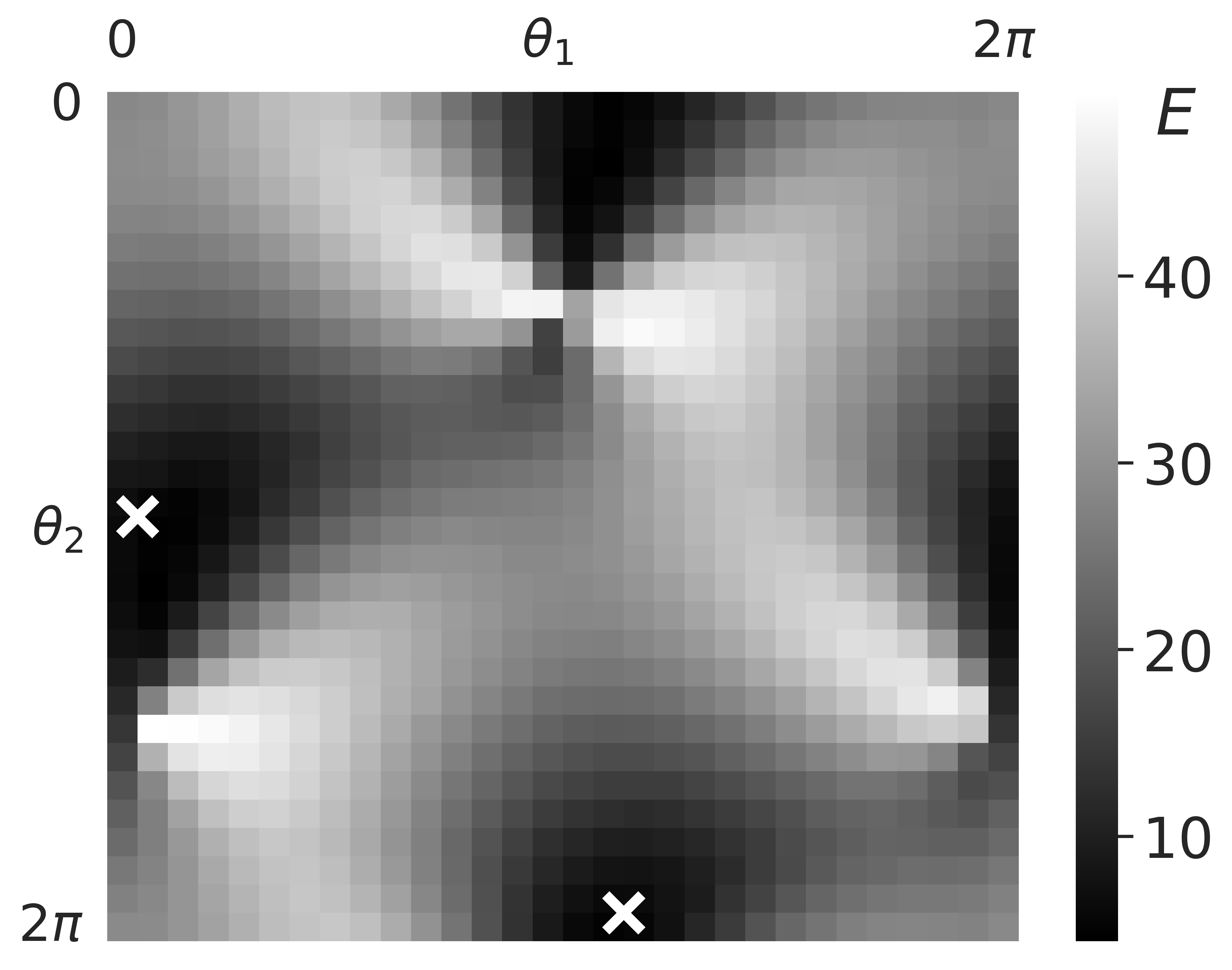}
\caption{\label{e_part1} The expectation value of the Hamiltonian $E=\bra{\phi(\vec{\theta})}H\ket{\phi(\vec{\theta})}$ as a function of two parameters $\theta_1,\theta_2$ for the random $\ket{b}$. Bright circles denote two equivalent minima of the function corresponding to the desired solution. The parameter space is specified on a 30 by 30 grid.}
\end{center}
\end{figure}

An example of the map of the Hamiltonian expectation value $E=\bra{\phi(\vec{\theta})}H\ket{\phi(\vec{\theta})}$ for two parameters $\theta_1,\theta_2$ is shown in Fig. ~\ref{e_part1} for the random $\ket{b}$. The bright dots indicate two solutions corresponding to $E=0$, which differ by modulo $\pi$.

The direct variational algorithm demonstrates a fundamental possibility to solve linear algebra problems, and in particular the discretized heat equation on a quantum processor. However it has one essential disadvantage: this algorithm does not demonstrate quantum speedup in the situation where the number of Pauli products in the Hamiltonian decomposition is exponential in the number of qubits. Generally, for the arbitrary $\ket{b}$, the number of terms in the Pauli products expansion is indeed exponential. In some cases it is possible to effectively sample over the Pauli products if one knows the distribution of the Pauli decomposition weights $c_i$, see Eq.(\ref{pauli_string_1}), but this option requires a separate study. Thus, its is more promising from a practical point of view to develop more sophisticated variational methods described below.

\section{Hadamard test based approach}
A development of the previous approach is a variational method where the problem of the measurement of the exponential (in general case) number of Pauli products is avoided. This method takes advantage of the fact that the Fourier mapping of the matrix $A$, given by Eq. (\ref{A_definition}), has a diagonal form:
\begin{eqnarray}
D=QFT^+A\,QFT, 
\end{eqnarray}
where $QFT$ is the quantum Fourier transform unitary matrix \cite{nielsen2002quantum}.

As in the preceding section we search for the minimum of the expectation value $E$ of the Hamiltonian (\ref{var_ham}). The expectation value $E =\langle x|H|x\rangle$ can be presented in the form,
\begin{eqnarray}
\begin{gathered}
E=\bra{x}A^+\left(I-\ket{b}\bra{b}\right)A\ket{x}=
\bra{x}A^+A\ket{x}-\left| \bra{x}A\ket{b}\right|^2=\\
=\bra{\phi}D^2\ket{\phi}-\left| \bra{\phi}D\ket{b_f}\right|^2=\\ 
=\bra{\phi}D^2\ket{\phi}-\Re\bra{\phi}D\ket{b_f}^2-\Im\bra{\phi}D\ket{b_f}^2,
\label{hadamarE}
\end{gathered}
\end{eqnarray}
where $\ket{\phi}$ is the Fourier transform of the desired solution and $\ket{b_f}$ is the Fourier transform of $\ket{b}$. For the matrix $A$ of the size $2^n \times 2^n $ this method requires $n+1$ qubits, where one extra qubit (ancilla) is exploited in the Hadamard test \cite{huang2021near}.

\begin{figure}
\begin{center}
\includegraphics[width=0.99\textwidth]{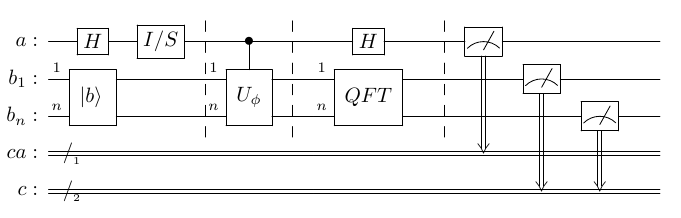}
\caption{\label{hadamar_test} A general quantum circuit of a variational algorithm using the Hadamard test. We choose gate $I$ or $S$ on the qubit $a$  to measure $\Re$ or $\Im$ $\left| \bra{\phi}D\ket{b_f}\right|$, respectively. To measure $\bra{\phi}D^2\ket{\phi}$, the $U_\phi$ gate is used without control from the qubit $a$. }
\end{center}
\end{figure}

The overall circuit of the algorithm is shown in Fig.~\ref{hadamar_test}. A prepared state $\ket{b}$ is an input for the quantum circuit. The desired gate and control of the ancilla qubit depends on which particular term in the last line of the expression (\ref{hadamarE}) is going to be measured. The measurement of the $\bra{\phi}D^2\ket{\phi}$ term does not require the usage of the ancilla qubit and $U_\phi$ control gate is disabled, while for the measurement of $\Re/\Im\bra{\phi}D\ket{b_f}$ term one takes $I/S$ gate on the ancilla
\begin{eqnarray}
\Re/\Im\{\bra{\phi}D\ket{b_f}\} = \bra{\xi_{I/S}}Z\otimes D\ket{\xi_{I/S}},
\label{Hadamard_test_formula}
\end{eqnarray}
where $\ket{\xi_{I/S}}$ is the output state of the circuit (see Fig.~\ref{hadamar_test}). The expression (\ref{Hadamard_test_formula}) is an implementation of the Hadamard test. The measurement outcomes obtained on the qubits $b_1 \dots b_n$ (see Fig.~\ref{hadamar_test}) is the binary representation of the eigenvalue index the operator $A$. The terms $\Re/\Im\bra{\phi}D\ket{b_f}$ are obtained by sampling out the eigenvalue index $\vec{b}=b_1 \dots b_n$ and the ancilla qubit value $a$ and averaging the corresponding eigenvalue with the weight factor $(-1)^{a}$. 

Thus, in order to measure the total expectation value of the Hamiltonian (\ref{hadamarE}) it is necessary to run only three quantum circuits and sampling out qubits outcomes $a,b_1,...,b_n$.
The variational parameters $\vec{\theta}$ enter into the circuits through the gate $U_\phi$:
\begin{eqnarray}
\begin{gathered}
\ket{\phi(\vec\theta)}=QFTU_\phi(\vec{\theta})\ket{b}.
\label{ansatz}
\end{gathered}
\end{eqnarray}

One searchers for the minimum of the expectation value $E$ of the Hamiltonian  (\ref{var_ham}) by varying the parameters $\vec\theta$. The obtained solution $\vec{\theta}_m$ gives the desired solution $\ket{x}$ using the $U_\phi$ ansatz:
$$
\ket{x}=U_\phi(\vec{\theta}_m)\ket{b}.
$$

\begin{figure}
\centering
\subcaptionbox{}{\includegraphics[width=0.76\textwidth]{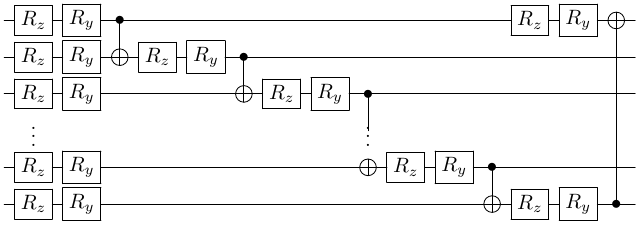}}%
\\ \vspace{2.0em}
\subcaptionbox{}{\includegraphics[width=0.50\textwidth]{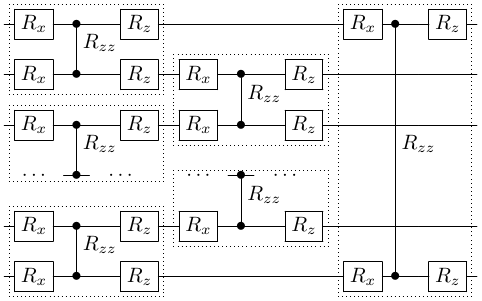}}%
\hfill 
\subcaptionbox{}{\includegraphics[width=0.40\textwidth]{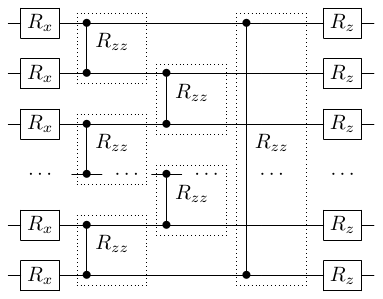}}%
\caption{\label{an_part2}\,One layer of the Hardware Efficient ansatz (a), the Checkerboard ansatz (b), Digital-Analog ansatz (c). The entangling gate (dotted region) is used in (b) and (c), where~$R_{zz}(\theta)=e^{-i\theta Z\otimes Z}$ and $\theta$ is a variational parameter. The number of gates in each layer is $\mathcal{O}(n)$, where $n$ is the number of qubits. }
\end{figure}

The main challenge of this method is a proper choice of the ansatz for $U_\phi$. The common recipe is to make the ansatz layered where the number of layers $M$ determines the precision. Fig.~\ref{an_part2} shows a single layer for each ansatz we use: (a) Hardware Efficient ansatz (HEA), (b) Checkerboard ansatz (CBA), and (c) Digital-Analog ansatz (DAA). DAA implies that a digital-analog strategy is used, which is based on always-on interaction between the qubits, so that single-qubit gates are implemented digitally, while the entanglement arises from native interactions of the qubits. The interaction of the qubits is assumed to be of Ising $ZZ$ type between the neighboring qubits of the quantum device, while the connectivity topology of the device is assumed to be a ring. All interaction constants are chosen to be equal to each other. Digital-analog strategy has an advantage that it does not require two-qubit gates. Note that quantum Fourier transform can also be implemented using the digital-analog approach \cite{we}. 

Each ansatz consists of the same layers where the number of the layers is determined by the problem. The number of gates in each ansatz is $\sim \mathcal{O}(n)$, where $n$ is the number of qubits. The quantum algorithm outperforms the classical one if the required number of the layers $M$ remains polynomial in $n$. In this situation the whole quantum algorithm will be completed in $\mathcal{O}(n^k) = \mathcal{O}(\log^k N)$ steps while the best classical algorithm for linear algebra problem requires $\mathcal{O}(N)$ steps, where $N=2^n$ is the matrix size.

In our work we use the measure of similarity of two vectors defined as
\begin{eqnarray}
F(a,b)=\left|\frac{a^\dagger b}{\sqrt{a^\dagger ab^\dagger b}}\right|^2.
\label{Fidelity_of_vectors}
\end{eqnarray}
This quantity turns into a well-known measure of the proximity of two pure quantum states if the vectors are normalized.

\begin{figure}[!ht]
\begin{center}
\includegraphics[width=0.99\textwidth]{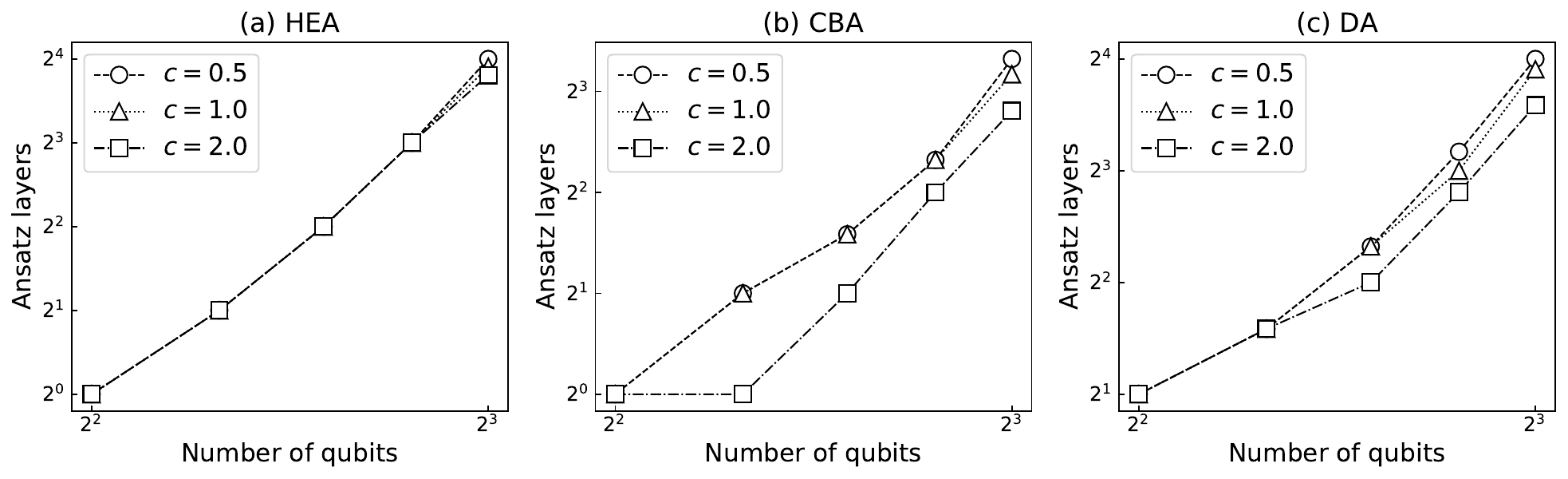}
\caption{\label{m_part2}The number of ansatz layers $M$  needed to achieve a fidelity of $0.99$, as a function of the number of qubits $n$ used. The graphs are plotted on the $\log-\log$ scale. A simulation was performed for three different types of ansatzes: HEA (a), CBA (b), and DAA (c). }
\end{center}
\end{figure}

The algorithm was run on a local simulator in the QISKIT package. For a given number of qubits $n$ and layers $M$ of the ansatz,  the optimization was carried out (finding the minimum of the expectation value of the Hamiltonian) and the fidelity (\ref{Fidelity_of_vectors}) between the resulting approximate and exact solutions was calculated. The number of the layers $M$ was increased until the fidelity value $F=0.99$ was achieved. The simulation results for the different ansatzes and parameters $c$ are shown in Fig.~\ref{m_part2}. Graphs are plotted on the $\log-\log$ scale, so any straight line indicates a polynomial dependence.

The simulation was carried out for the number of qubits $n$ ranging from 2 to 8. Each point on the graph was obtained by averaging out the results for a random sample of 20 vectors $\ket{b}$. It was found that the number of ansatz layers $M$ required to achieve a given accuracy is practically independent on the initial conditions (vector $\ket{b}$).

The curves in Fig.~\ref{m_part2} show superlinear behavior in most cases.  At the same time, in the case of CBA (b) for the parameter $c=2$ we see a polynomial dependence and the number of layers $M$ needed to solve the problem remains polynomial in the number of qubits. The best results were obtained for CBA, as it provides a more uniform entanglement. Moreover, an increase in the parameter $c$ also leads to a decrease in the number of the required ansatz layers. The weak convergence of the variational algorithm for small $c$ can be explained by the concept of condition number, which is outlined in the Section II. However, the present simulations with at most 8 qubits turn out to be not sufficient to make an unambiguous conclusion about the advantage of the quantum algorithm over the classical one. Still we can conclude that it is important to choose both an appropriate ansatz and the value of parameter $c$.  In the next section we demonstrate how to choose the ansatz which provides a computational advantage.




\section{Ansatz tree approach}

In this section, we apply the variational algorithm of Ref. \cite{huang2021near} to the solution of the heat equation. The algorithm has a branched tree structure and is referred to as the ansatz tree approach (ATA). The ATA is based on an efficient decomposition of the matrix $A$ (\ref{A_definition}) into a superposition of unitary matrices
\begin{eqnarray}
A=\sum_{i=1}^{K_A}\beta_iU_i.
\label{A_ansatz_tree_decomposition}
\end{eqnarray}
The efficient decomposition means that: (i) the unitary operators of the decomposition can be presented by quantum circuits with a polylogarithmical depth, (ii) the number of $U_i$ also scales polylogarithmically with the size of the matrix $A$. The loss function of this approach has the form
\begin{eqnarray}
L_R(x)=\| Ax-\ket{b}\|^2_2= x^\dagger A^\dagger A x - 2 Re\{ x^\dagger A\ket{b} \} + 1;
\label{Ansatz_tree_approach_loss_function}
\end{eqnarray}
while the gradient overlap is defined as
\begin{eqnarray}
\nabla L_R(x)=2A^2x-2A\ket{b}.
\label{Ansatz_tree_approach_loss_function_gradient_overlap}
\end{eqnarray}
The loss function is the well-known $\ell_2$-norm loss used in regression methods. One looks for the solution vector $x$ in the tree form 
\begin{eqnarray}
x=\alpha_0\ket{b}+\alpha_1U_{v_1}\ket{b}+\alpha_2U_{v_2}U_{v_1}\ket{b}+\dots,
\label{x_ansatz_tree_approach}
\end{eqnarray}
where $\alpha_k$ are variational parameters. The parameters $v_j$ determine the index of $U_i$ from the decomposition (\ref{A_ansatz_tree_decomposition}). The rule of expansion of the vector $x$ is explained below. Note that the vector $x$ is not normalized as far as the vectors $\ket{b}$, $U_{v_1}\ket{b},$ $U_{v_2}U_{v_1}\ket{b}\dots$ are not orthogonal. Reference \cite{huang2021near} proves that the loss function (\ref{Ansatz_tree_approach_loss_function}) is convex when $x$ has the form (\ref{x_ansatz_tree_approach}). For the further convenience, we introduce the vectors
\begin{eqnarray}
\ket{0}=\ket{b},\qquad \ket{1}=U_{v_1}\ket{b},\qquad \ket{j}=U_{v_j}U_{v_{j-1}}\dots U_{v_1}\ket{b},
\label{labeled_states_ansatz_tree_approach}
\end{eqnarray}
\begin{eqnarray}
x=\sum_j\alpha_j\ket{j}.
\label{x_through_labeled_states_ansatz_tree_approach}
\end{eqnarray}

Let us consider an iterative optimization algorithm for the expansion of the vector $x$. In the beginning, let the subspace $S$ contains only the root of the ansatz tree, i.e., $S = \{\ket{b}\}$. At each next step, we perform the following:
\begin{enumerate}
\item Find the optimal $x^s=\sum_{j=0}^m\alpha_j\ket{j}$ by optimizing the loss function (\ref{Ansatz_tree_approach_loss_function}) over the parameters $\alpha_0,\dots,\alpha_m$.

\item For each child quantum state $\ket{c} \in C(S) =\{U_1\ket{m},U_2\ket{m},\dots,U_{K_A}\ket{m}\} $ compute the gradient overlap $g=\bra{c}\nabla L_R\left(x^s\right)=2\sum_{j=0}^m\alpha_j\bra{c}A^2\ket{j}-2\bra{c}A\ket{0}$. This quantity can be computed efficiently using the quantum circuit depicted in Fig. \ref{hadamar_test}.
\item  Add a new node with the largest gradient overlap to the subspace $S$:  $S\leftarrow S\cup \{\ket{m+1}\}$,  $\ket{m+1}=\mbox{argmax}_{\ket{c}\in C\left(S\right)}|g|$.
\end{enumerate}

\subsection{Efficient unitary decomposition for heat equation}

In this subsection we apply ATA to the heat equation. First, we construct the efficient unitary decomposition of $A$. Obviously, any hermitian matrix can be decomposed into the Pauli products. This decomposition, indeed, satisfies the first efficiency requirement, but scales exponentially with the number of qubits. Nevertheless, we demonstrate that the decomposition into the Pauli products can be effectively incorporated into our approach. Our idea is to switch into the Fourier representation and modify the matrix $A$ spectrum, that will sufficiently reduce the number of the decomposition terms. 

\subsubsection{Piecewise quadratic approximation of the spectrum for the heat equation}

\begin{figure}[ht]
\includegraphics[width=0.6\textwidth]{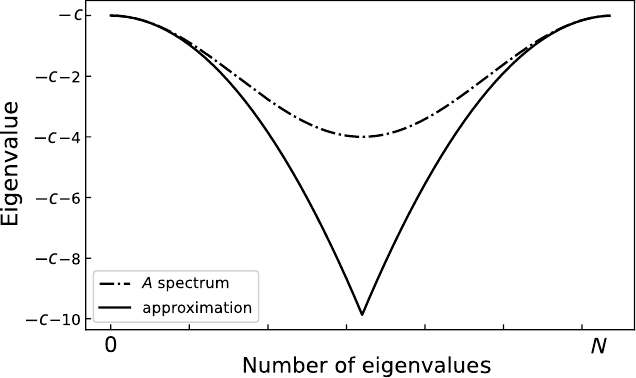}
\caption{The spectrum of the matrix $A$ and its piecewise quadratic approximation at small $\lambda^\Delta_k$.}
\label{Spectrum_and_substituted_spectrum}
\end{figure}

Let us consider the spectrum of the matrix (\ref{A_definition})
\begin{eqnarray}
\lambda_k^A=\left(QFT\, A\, QFT^\dagger\right)_{kk}=\left(A_F\right)_{kk}=-c-4\sin^2\left(\frac{\pi k}{2^n}\right),
\label{A_spectrum}
\end{eqnarray}
where $QFT$ is the Fouruer transform unitary transformation matrix. The Laplace operator $\Delta$ has a quadratic spectrum $\Delta\phi_k=k^2\phi_k$, where $\phi_k$ are eigenvectors. The finite difference representation of the Laplace operator, see Eq.(\ref{Thermal_conductivity_equation_grid_initial_decomposition}), modifies the quadratic spectrum into the sine spectrum (\ref{A_spectrum}). The eigenvalues of these two spectra coincide at small $k$. Therefore, we replace the spectrum of the matrix $A$ by a piecewise quadratic spectrum for small $\lambda^\Delta_k=k^2$, see Fig. \ref{Spectrum_and_substituted_spectrum}. This transformation can be done in the following way: first, we apply $N/2$ times the cyclic permutation to the diagonal matrix $\left(A_F\right)_{kk}$, 
\begin{eqnarray}
\left(A_F\right)_{kk} \to \left(A^L_F\right)_{kk}=-c-4\sin^2\left(\frac{\pi k}{2^n}- \frac{\pi}2\right),
\label{A_fourier_diag_shifted}
\end{eqnarray}
and, next, we expand the sine at the maximum:
\begin{eqnarray}
\left(A^L_F\right)_{kk} \to \left(A^{L\prime}_F\right)_{kk}=-c-4\left(\frac{\pi k}{2^n}-\frac{\pi}{2}\right)^2.
\label{A_fourier_diag_substituted_spectrum_shifted}
\end{eqnarray}
Finally, we again apply the $N/2$-fold cyclic permutation,
\begin{eqnarray}
\left(A^{L\prime}_F\right)_{kk} \to \left(A^{\prime}_F\right)_{kk}=-c-\pi^2\left(|\frac{k}{2^{n-1}}-1|-1\right)^2.
\label{A_fourier_diag_substituted_spectrum}
\end{eqnarray}

This approximation implies that a sufficiently fine spatial grid is selected, so that the heat source  function $f(x,t)$ changes smoothly on the grid scale. 

\begin{figure}[t]
\includegraphics[width=0.6\textwidth]{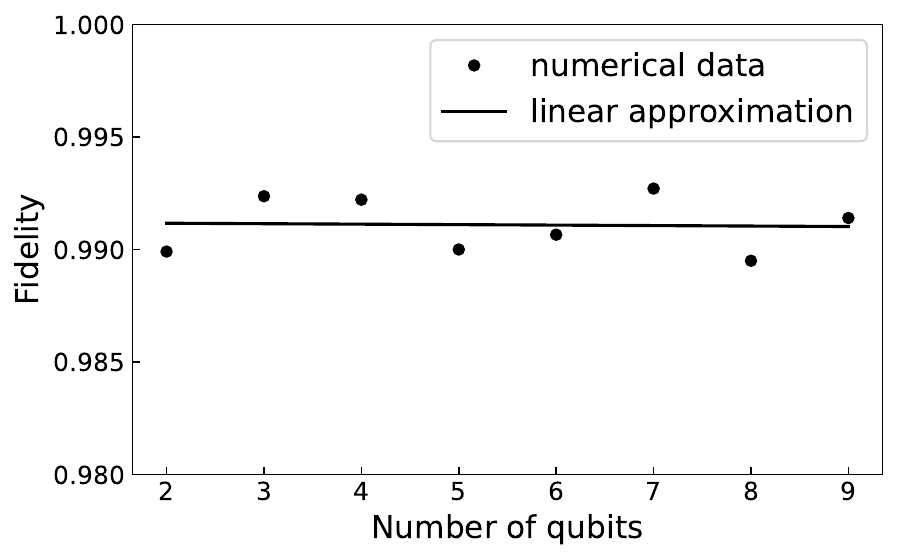}
\caption{The dependence of the fidelity of two solutions for the matrix $A$ with a sinusoidal and $A^\prime$ with a piecewise quadratic spectrum on the number of qubits (the size of the matrix $A$). Each point represents an averaging over 1000 random $\ket{b}$ and over 20 random values of $c$ in the range from 0.1 to 2.0}
\label{Approximation_efficiency}
\end{figure}

Let us analyze how the solution of the equation (\ref{Ax=b}) changes due to the approximation has been made.  Figure \ref{Approximation_efficiency} shows how fidelity between $A^{-1}\ket{b}$ and $A^{\prime-1}\ket{b}$ depends on the size of the matrix (the number of qubits). This figure evidences that the approximation of the spectrum (\ref{A_fourier_diag_substituted_spectrum}) introduces only a small perturbation to the solution, and the error does not scale with an increase of the number of qubits (the size of the system). Infidelity can be further reduced by eliminating high-frequency harmonics of Laplace operator. Thus, instead of the system (\ref{Ax=b}) with matrix (\ref{A_definition}), in the present paper the ATA is applied to the same system with the matrix $A^\prime$ without a significant loss of accuracy.

\subsubsection{Decomposition into Pauli products in the Fourier representation}

In this section, we construct the efficient decomposition of the matrix $A^\prime$ into a sum of unitary operators that satisfies both efficiency requirements. This decomposition is based on the statement: for a diagonal matrix with the spectrum $p(m) =\sum _{i=0}^s\alpha_im^i$, the decomposition into Pauli products involves only terms containing no more than $s$ operators $Z$. This statement is analyzed and proved in Appendix A. The matrix $A^\prime$ has the same eigenvectors as $A$, and therefore is diagonal in the Fourier representation. In this case, $A^\prime$ has the piecewise quadratic spectrum, which also satisfy the above statement. Hence the matrix $A^\prime$ can be decomposed into $\mathcal{O}(n^2)$ of Pauli products, each of which contains no more than two operators $Z$
\begin{eqnarray}
\begin{gathered}
A^{\prime}=QFT^\dagger\left(
\sum_{i,j}d_{ij}Z_iZ_j+\sum_is_iZ_i+\zeta I
\right)QFT=\\
\sum_{i\neq j}d_{ij}QFT^\dagger Z_iZ_jQFT+\sum_is_iQFT^\dagger Z_iQFT+\zeta I
\label{A_substituted_decomposition}
\end{gathered}
\end{eqnarray}

Thus, combinations of $Z_iZ_j$ and $Z_i$, as well as $I$  can be chosen as $U_i$ for the decomposition (\ref{A_ansatz_tree_decomposition}) in the Fourier representation. This satisfies the second efficiency requirement: the number of $U_i$ depends polylogarithmically on the size of the matrix $A^\prime$.

Let us give an upper estimate for the complexity of constructing of the unitaries $U_i$. The quantum Fourier transform requires $n(n-1)/2$ Cphase gates. Two of them are required, therefore, the construction of one $U_i$ requires at most $n(n-1)$ two-qubit entangling gates, and on average $n(n-1)/2$, since the quantum Fourier transformation collapses when applying $Z$ gates not to the last qubit. Thus, the unitary decomposition (\ref{A_substituted_decomposition}) satisfies both efficiency requirements.

\subsection{Numerical realization of the ansatz tree approach}

In this subsection, a numerical simulation of the described algorithm for the heat equation is considered. To implement the algorithm, it is necessary to be able to calculate the loss function (\ref{Ansatz_tree_approach_loss_function}) and the gradient overlap
\begin{eqnarray}
g=\bra{c}\nabla L_R\left(x^s\right)=2\sum_{j=0}^m\alpha_j\bra{c}A^2\ket{j}-2\bra{c}A\ket{0}.
\label{gradient_overlap_Ansatz_tree_approach}
\end{eqnarray}

\begin{figure}[t]
\begin{center}
\includegraphics[width=0.4\linewidth]{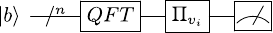}    
\end{center}
\caption{Quantum circuit for measuring  $\bra{i}A\ket{i}$ and $\bra{i}A^2\ket{i}$.}
\label{mean_measure_ansatz_tree}
\end{figure}

\begin{figure}[ht]
\begin{center}
\includegraphics[width=0.6\linewidth]{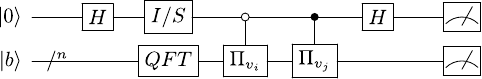}    
\end{center}
\caption{Quantum circuit for measuring $\Re/\Im\{\bra{i}A\ket{j}\}$ and $\Re/\Im\{\bra{i}A^2\ket{j}\}$ depending on the $I/S$ gate.}
\label{ij_hadamard_measurement_circuit}
\end{figure}

A quantum computer is used to calculate the values of $\bra{i}A\ket{j}$ and $\bra{i}A^2\ket{j}$. Similarly to the Section 4, the Hadamard test is utilized, where the corresponding quantum circuits in the ATA case are shown in Fig.~\ref{mean_measure_ansatz_tree} and \ref{ij_hadamard_measurement_circuit}. Here we exploit the decomposition of the matrix $A$ into the Pauli products $\Pi_{v_i}$ in the Fourier representation and denote
\begin{eqnarray}
\ket{i}  = U_{v_i}\ket{b} = QFT^\dagger\Pi_{v_i} QFT\ket{b}.
\label{Pauli_string_PI_definition}
\end{eqnarray}
The measurement outcome is the eigenvalue index of the matrix $A$ or $A^2$, respectively, similar to the method described in the Section 4.

In contrast to the methods described in the Sections 3 and 4, the ATA does not require a permanent access to the quantum device during the optimization of the loss function. Here, the optimization occurs in two separate steps: i)  quantum measurements of all terms $\bra{i} A^2 \ket{j}$ and $Re \bra{i} A \ket{0}$ in the decomposition (\ref{x_ansatz_tree_approach}) by the Hadamard test; and ii) classical optimization of the loss function $L_R(x)$ with respect to the parameters $\alpha_i$. Moreover, when extending the tree depth from $d$ to $d+1$, only the new unknown terms $\bra{i}A^2\ket{d+1}$, $i=0,\dots,d$ and  $Re \bra{d+1}A\ket{0}$ must be measured. On the contrary, the optimization of the variational parameters $\vec\theta$ of the ansatz circuit $U(\vec\theta)$ requires the new measurement of the loss function for each new change in $\vec\theta$.

In Appendix B, we discuss the influence of the depolarizing noise on this type of optimization. In particular, we found that in the case of a fully depolarizing channel the solution accuracy does not converge to $1/2^n$, which corresponds to a random vector.

\subsubsection{Construction of the solution}

\begin{figure}[!ht]
\includegraphics[width=0.6\linewidth]{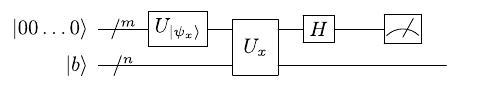}
\caption{The quantum circuit which generates the normalized solution $\ket{x}$.}
\label{x_preparation_ansatz_tree_approach_circuit}
\end{figure}

Given the set of weights $\{\alpha_i\}$, the quantum solution $\ket{x}$ can be constructed in a probabilistic way by the circuit  shown in Fig.~\ref{x_preparation_ansatz_tree_approach_circuit}.  The quantum gate $U_{\ket{\psi_m}}$ generates the state $\ket{\psi_m}=U_{\ket{\psi_m}} \ket{00\dots 0}$ in the upper auxiliary qubit register, where 
\begin{eqnarray}
\ket{\psi_x}=\sum_{i=0}^{m-1}\alpha_i\ket{i}.
\label{U_psi_x_ansatz_tree_approach}
\end{eqnarray}
The gate $U_x$ is defined as,
\begin{eqnarray}
U_x=\sum_{j=0}\ketbra{j}{j}\otimes U_{i_j}U_{i_{j-1}}\dots U_{i_1},
\label{U_x_ansatz_tree_approach}
\end{eqnarray}
and can be constructed through two Fourier transforms and $CZ$ gates in between. The solution $\ket{x}$ is generated in the lower qubit register once one measures the state $\ket{00\dots 0}$ in the auxiliary register. The number of the auxiliary qubits scales as $\lceil\log_2d\rceil$ with the tree depth $d$. Thus, on average, the probability of triggering the solution $\ket{x}$ is $1/2^{\lceil \log_2d\rceil}$.

\subsubsection{Estimation of algorithm complexity}

Let us estimate the complexity of the ATA - the total number of single and two-qubit gates applied during the construction of the ansatz tree of the depth $d$. First, let us find the number of operations applied for the loss function measurement, see Eq.(\ref{Ansatz_tree_approach_loss_function}). At each step $m=0,\dots d-1$ one needs to perform $m+2$ separate measurements: $m+1$ measurements $\bra{m}A^2\ket{i}$, $i=0,\dots,m$ and one $Re \bra{m} A \ket{0}$ measurement. Each of these measurements is done with Hadamard test, see Fig. \ref{ij_hadamard_measurement_circuit}, involving $\mathcal{O}(n^2)$ quantum gates due to Fourier transform circuit. The net cost of the loss function measurement is thus given by $\mathcal{O}(d^2 n^2)$. Second, during the tree growth one searches $d$ times the best candidate for the next level of the tree among $n^2$ candidates. For each candidate one estimates the overlap gradient, see Eq.(\ref{gradient_overlap_Ansatz_tree_approach}), that requires $d$ measurements of the Hadamard test with $\mathcal{O}(n^2)$ quantum gates circuits. Thus in total the whole  gradient overlap procedure exploits $\mathcal{O}(d^2 n^4)$ quantum gates and the net complexity of the ATA is $\sim \mathcal{O}(d^2n^2+d^2n^4)\sim\mathcal{O}(d^2n^4)$. The classical tridiagonal matrix inversion algorithm \cite{lee2011tridiagonal} has complexity $\mathcal{O}(2^n)$ and, therefore, the ATA gives the exponential speed-up compared to the classical algorithm at $d=poly(n)$. 




The Pauli decomposition of the inverse matrix $(A^\prime)^{-1}$ has in general $2^n$ terms,
\begin{eqnarray}
A^{\prime-1}_F=\sum^{2^n-1}_{p=0}h_p\Pi_p,
\label{A_inv_decomposition_to_pauli_products_fourier_space_ansatz_tree_approach}
\end{eqnarray}
where the weights $h_p$ can be expressed through the weighted sum of the eigenvalues of the matrix $A^{\prime-1}$,
\begin{eqnarray}
h_p=\sum_{i=0}^{N-1}\lambda^{A^{\prime-1}}_i(-1)^{\sum_si_sp_s},
\label{coeeficients_of_decomposition_to_Pauli_products_A_inv_ansatz_tree_approach}
\end{eqnarray}
where $p_s$ and $i_s$ are the $s$th binary digit of $p$ and $i$ index, respectively. 

\begin{figure}[t!]
\includegraphics[width=1\linewidth]{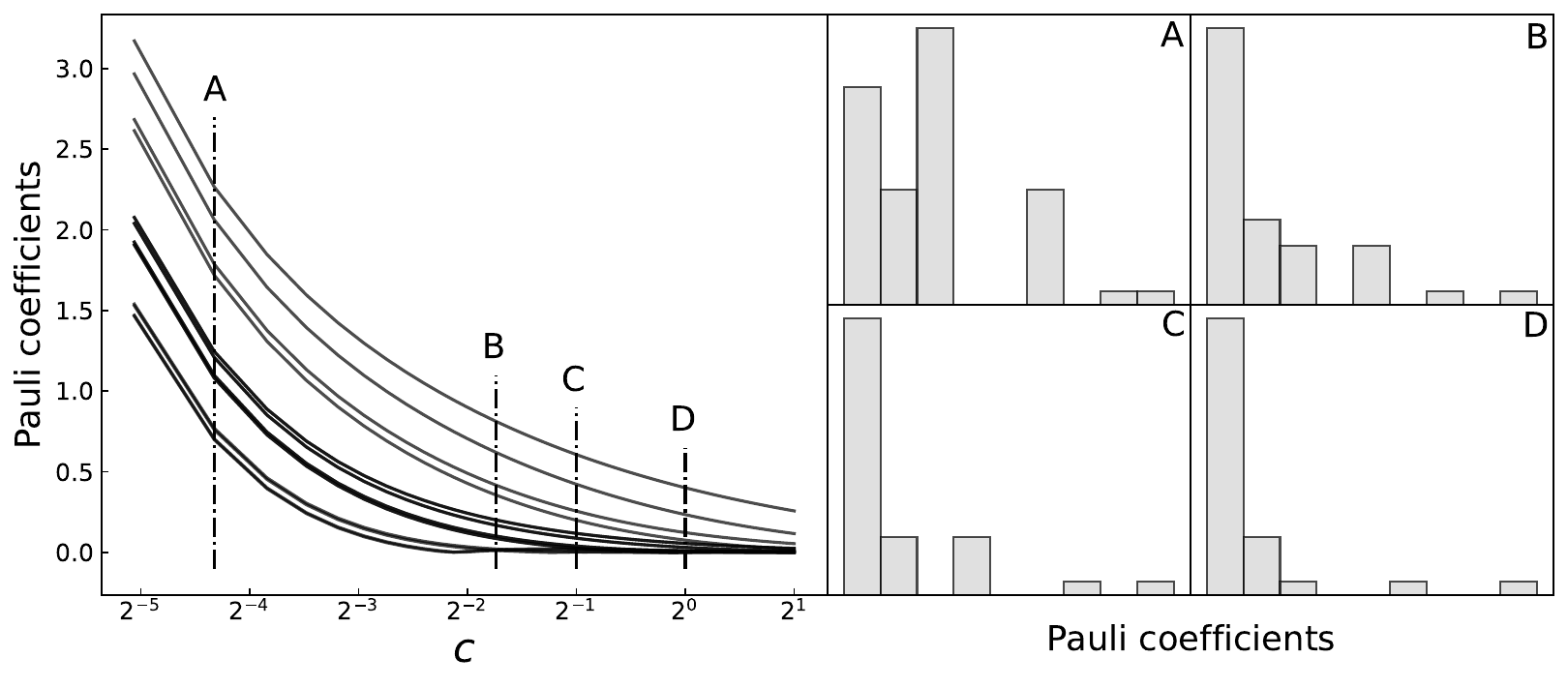}
\caption{The absolute value of nonzero coefficients of the decomposition into Pauli products of matrix $A^{\prime-1}_F$ as a function of $c$ for 4 qubits. On the right there are the Pauli coefficients distributions for $c$ from the set \{0.05, 0.3, 0.5,1\} for 6 qubits. Distribution histograms have a logarithmic scale along the vertical axis. }
\label{Pauli_coefficient_of_inverted_matrix}
\end{figure}

For the matrix $A^{\prime-1}$, the analytic form $h_p$ is cumbersome and therefore is not amenable to an explicit analysis. Fig. \ref{Pauli_coefficient_of_inverted_matrix} shows the dependence of non-zero $\left|h_p\right|$ on the parameter $c$ (\ref{c_definition}). One can see, as far as the parameter $c$ growths the distribution of $\left|h_p\right|$ shrinks near zero values and only few non-vanishing $\left|h_p\right|$ contribute to $A^{\prime-1}_F$. We assume that this behaviour on the parameter $c$ is preserved with rising size of the matrix $A$. The ATA at each step selects a Pauli product with the largest in absolute value  weight. Thus our conjecture is that the depth of the ATA tree can be upper bounded by the $d \leq poly(n)$, and the contribution of the Pauli products with a relatively small $\left|h_p\right|$ can be neglected without significant loss in the accuracy of the ATA solution,
\begin{eqnarray}
x= A_{sol}^{ATA}\ket{b}=(\alpha_0I+\alpha_1U_{i_1}+\alpha_2U_{i_2}U_{i_1}+\dots)\ket{b}.
\label{A__inv_in_fact_ansatz_tree_approach}
\end{eqnarray}
We note, that the matrix $A_{sol}^{ATA}$ is build for the specific $\ket{b}$ and thus is not in general the approximation of the inverse matrix $A^{\prime-1}$. 

\begin{figure}[ht]
\includegraphics[width=0.6\linewidth]{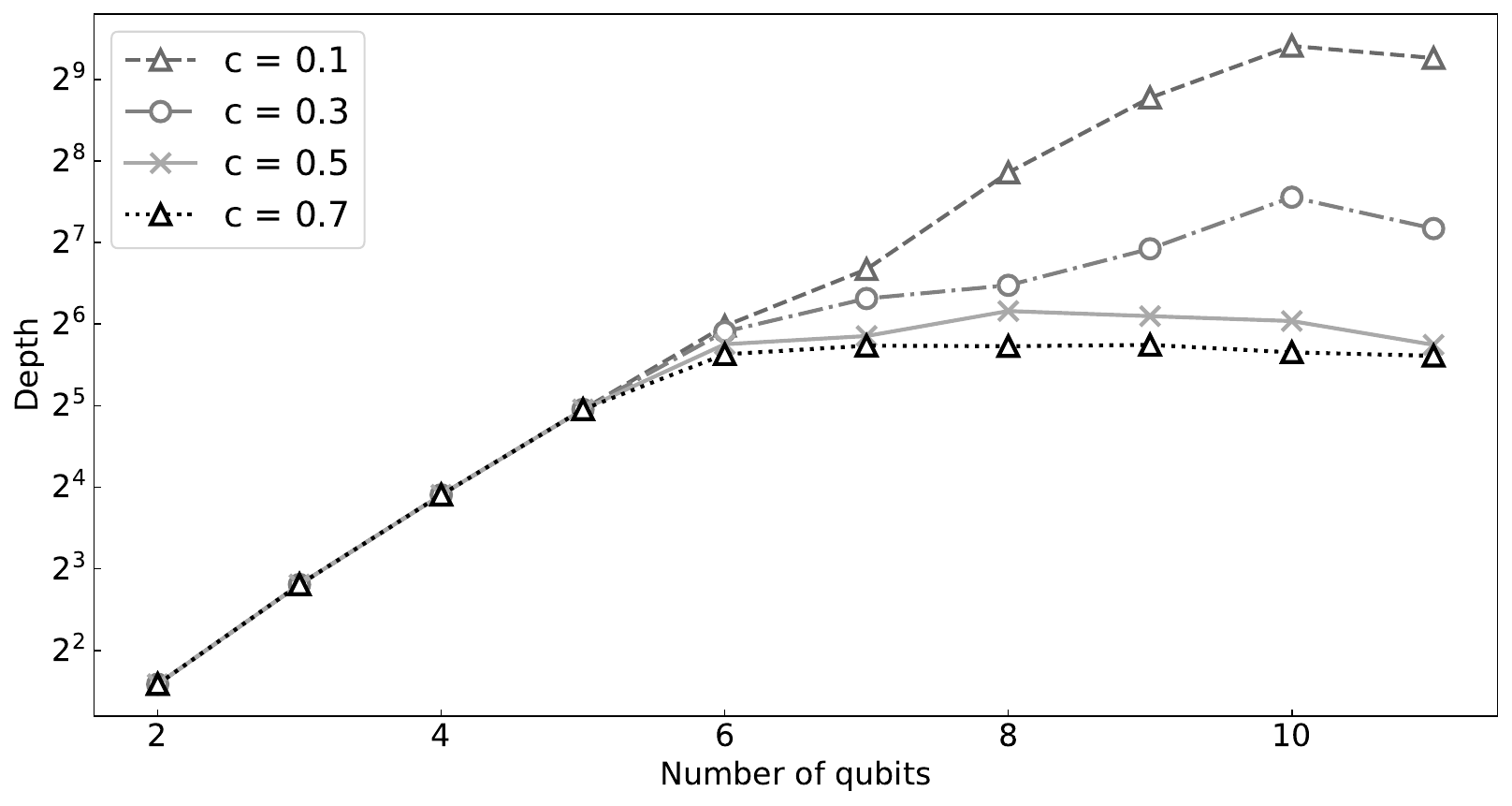}
\caption{Dependence of the depth of the tree reaching the fidelity of 0.99 on the number of qubits (size of the matrix $A$) for the different values of the grid parameter $c$.}
\label{dependence_depth_n_qubits_fidelity99_ansatz_tree_approach}
\end{figure}

Therefore, we solve the linear system  (\ref{Ax=b}) with a finite accuracy. In order to analyze the convergence of the ATA, we limit the expansion of the ansatz tree by the fidelity value 0.99  between the ATA solution and the exact numerical solution for matrix $A^\prime$. Figure \ref{dependence_depth_n_qubits_fidelity99_ansatz_tree_approach} shows the dependence of the depth corresponding to the fidelity value 0.99 on the number of qubits (matrix size $A$). One can see, that as the parameter $c$ increases, the exponential dependence turns into a constant one in accordance with our conjecture.

The parameter $c$ is controllable through the grid sizes $\delta z$ and $\delta t$, see   (\ref{c_definition}). However, choosing a very large parameter $c$ gives too tiny step in the temporal dimension and thus may lead to a loss in the computation speed. Thus, based on the results of numerical simulation, we can conclude that with the correct choice of the grid parameter $c$, it is possible to achieve a depth that is polynomial or even constant in the number of qubits and reaches a given accuracy. It follows from the estimate $\mathcal{O}(d^2n^4)$ that, for the polynomial depth, ATA exponentially outperforms the classical tridiagonal matrix algorithm whose complexity is estimated as $\mathcal{O}(2^n)$.

\subsection{Generalization to higher dimensions}

Hitherto we have considered the case of the heat equation with one spatial dimension. The generalization to the multidimensional case is straightforward. The multidimensional analog of the matrix $A$ has the form,
\begin{eqnarray}
A^{\left(d_r\right)}\left(c\right)=\underbrace{A\left(0\right)\otimes I\otimes I\otimes\dots\otimes I}_{d_r}+I\otimes A\left(0\right)\otimes I\otimes\dots\otimes I+\dots+I\otimes I\otimes I\otimes\dots\otimes A\left(0\right)-cI^{\left(2^nd_r\right)}.
\label{A_coordinate_scaling}
\end{eqnarray}
This form of the matrix $A^{(d_r)}$ allows us to determine the effective set of $U_i$ for the ATA algorithm. For example, the $i$th term of the sum (\ref{A_coordinate_scaling}) can be decomposed as,
\begin{eqnarray}
\underbrace{I\otimes I\otimes I\otimes\dots\otimes I}_{i-1}\otimes A\otimes\dots\otimes I=\underbrace{I\otimes I\otimes I\otimes\dots\otimes I}_{i-1}\otimes \sum_{i=1}^{K_A}\beta_iU_i\otimes\dots\otimes I,
\label{i_A_coordinate_scaling}
\end{eqnarray}
where the decomposition of the matrix A (\ref{A_ansatz_tree_decomposition}) has been used. Hence, the ATA scales as {$\mathcal{O}(d_rd^2n^4)$} depending on the dimension of the coordinate space $d_r$, where $n$ is the number of qubit required for a single spatial dimension.

\section{Analysis of the ansatz tree approach complexity}

In this section, we compare the performance of the Ansatz-Tree approach and the HHL algorithm \cite{hhl} in solving the heat equation. We start our consideration with a fault-tolerance case. The HHL algorithm is probabilistic; it returns the solution when its ancilla qubit is measured in the state $|1\rangle$ with some probability $p<1$. The solution of the heat equation (\ref{Thermal_conductivity_equation}) requires $N_\tau$ times applications of the HHL algorithm to the linear system (\ref{equation_grid}) at each time step, where the initial condition at the $i$th time step involves the solution of the preceding $i\!-\!1$th time step, see Eq.(\ref{c_definition}). This means that the correct solution of the heat equation at the final time step is obtained only with an exponentially small probability $\sim (1-p)^{N_\tau} \approx \exp(-pN_\tau)$ implying that the ancilla qubit is measured in the state $|1\rangle$ at the all sequential applications of the HHL algorithm. Thus the probabilistic nature of the HHL algorithm significantly restricts its application for the heat equation problem even in the fault-tolerant case.

In the ansatz tree approach one also generates the solution of the heat equation in the probabilistic manner as shown in the Fig. \ref{x_preparation_ansatz_tree_approach_circuit}, 
where the solution is returned only when $m$ ancilla qubits are all measured in the state $|0\rangle$ with the probability $\sim 1/d$ where $d$ is the anzatz depth. The important difference with the HHL approach is that we build up the ansatz tree in a classical manner: once we learned the ansatz tree for the time step $t$ we need not repeat all the proceeding steps while constructing the initial condition for the next time step. This allows us to circumvent the exponentially decay of the probability to get the correct answer: the probabilistic overhead in the number of runs of the circuit Fig. \ref{x_preparation_ansatz_tree_approach_circuit} results in  the penalty factor $\sim N_\tau d$ which scales linearly with $N_\tau$ in contrast to the exponential penatly factor $\exp(N_\tau p)$ in the number of runs of the HHL algorithm.  

The construction of the initial condition for the next time step can be done efficiently, provided the heat source function $\vec{f}^t$ can be generated at each time by the known ansatz tree applied to the initail state $\vec{b}^0$:
\begin{eqnarray}
\vec{f^t}=QFT^\dagger\left(\sum^{K^t_f}_{i=0}\beta_i^t\Pi_{h_i}\right)QFT\, \vec{b}^0,
\label{heat_sources_special_view}
\end{eqnarray}
where $\Pi_{h_i}$ is a Pauli product that contains tensor products of $X$, $Y$, $Z$, $I$.

Indeed, as follows from the Eqs. (\ref{c_definition}), (\ref{x_ansatz_tree_approach}), (\ref{heat_sources_special_view}) the initial condition for the time step $t+1$ is given by,
\begin{eqnarray}
\begin{gathered}
\vec{b}^t=\frac{(\delta z)^2}{a^2}\vec{f}^t+c\,\vec{x}^t=\frac{(\delta z)^2}{a^2}QFT^\dagger\left(\sum^{K^t_f}_{i=0}\beta_i^t\Pi_{h_i}\right)QFT\vec{b^0}+c\,QFT^\dagger\left(\sum^{K^t_x}_{i=0}\alpha_i^t\Pi_{v_i}\right)QFT\,\vec{b}^{t-1}=\dots=\\QFT^\dagger\left(\sum^{K_b^0}_{i=0}\kappa^0_i\Pi_{s_i}\right)QFT\,\vec{b^0}.
\label{initial_cond_ith_forming}
\end{gathered}
\end{eqnarray} 

Next, let us discuss the noise sensitivity of the variational ansatz tree approach and the HHL approach due to the gate errors. The most important parameter in this respect is the number of two qubits gates required to build the variational ansatz, see Fig. \ref{x_preparation_ansatz_tree_approach_circuit}, and the HHL quantum circuit. In the former case the majority of the two-qubit gates come from the Fourier transform which results in $O(n^2)$ two-qubit gates complexity of the variational ansatz. The HHL quantum circuit comprises three different blocks: i) a phase estimation block (PEA), ii) a reciprocal eigenvalue block (REV) and iii) an inverse phase estimation block (iPEA). The PEA block decomposes the initial vector $\vec{b}^0$ into eigensystem of the square matrix $A$, see Eq.(\ref{A_definition}): $|0\rangle |\vec{b}^0\rangle \to \sum_\lambda \beta_\lambda^0 |\lambda\rangle |\vec{a}_\lambda\rangle$ with $A \vec{a}_\lambda = \lambda \vec{a}_\lambda$. For the specific heat equation case the PEA and iPEA blocks can be implemented with $O(n^3)$ number of two-qubit gates. The IEV block makes the reciprocal function transformation of the eigenvalue register and the ancilla qubit: $|0\rangle |\lambda\rangle \to (\sqrt{1-c^2/\lambda^2} |0\rangle + c/\lambda |1\rangle)|\lambda\rangle$. In general, for an arbitrary spectrum $\{\lambda\}$ this transformation requires $O(\exp(n))$ quantum gates, although for the quadratic spectrum, see Eq.(\ref{A_fourier_diag_substituted_spectrum}), one possibly can compose an approximate REV operator with polylog complexity $O(n^p)$ that requires a separate study. Therefore, one can conclude that HHL approach requires at least $O(n^3)$ two-qubit gates that makes it more sensitive to the gate noise.

Now we turn to the complexity of the construction of the initial state $\vec{b
}^0$ which was not considered in the previous sections. The construction of an arbitrary state vector is an exponentially hard problem in general. In order to have quantum advantage one has conjecture an efficient way for generation of the initial state $\vec{b}^0$. For the ansatz tree approach considered here any generation procedure which has $O(poly(n))$ complexity guarantees an exponential speedup over the classical algorithm, see the part B.2 of the Section V. We argue, that such a requirement corresponds a sufficiently smooth initial temperature distribution function \cite{grover2002creating}. We interpret this physically meaningful restriction as a limit on the number of non-zero low-frequency harmonics in the Fourier decomposition of the initial temperature distribution.

\section{Analysis of the algorithms accuracy}

In this Section, we evaluate the errors inherent in the variational algorithms discussed in this paper. The main parameter under study is the grid parameter $c$ of the finite difference scheme, which controls the number of time partitions $N_\tau$ for the fixed number of coordinate partitions $N_z$.  As mentioned earlier, the number $N_z$ is determined by the number of qubits. 
At each time step $i=1,\dots N_\tau$ we define the variational algorithm error $\epsilon_i$ as the infidelity between the solution $x_i$ given by the classical numerical algorithm and the solution $\tilde{x}_i$ given by the variational algorithm,
\begin{eqnarray}
\epsilon_i = 1- |x_i^\dagger\widetilde{x_i}|^2.
\label{Error_fidelity_definition}
\end{eqnarray}
Then it follows from (\ref{Error_fidelity_definition}) that
\begin{eqnarray}
|x_i^\dagger\delta x_i|=\epsilon_i/2+o(\epsilon_i^2),
\label{Error_fidelity_xdx}
\end{eqnarray}
where $\delta x_i \equiv \tilde{x}_i-x_i$. If the number of partitions $N_\tau$ is large enough, one can assume that $x_i$ is almost the same as $x_{i+1}$ and therefore  
\begin{eqnarray}
|x_{i+1}^\dagger\delta x_i|\approx|x_i^\dagger\delta x_i|=\epsilon_i/2+o(\epsilon_i^2).
\label{Error_fidelity_Assumption}
\end{eqnarray}
According to the Eq.(\ref{c_definition}) the solution from the previous time instant $\tilde{x}_i$ enters to the next time step of the variational scheme through the vector $\tilde{b}_i = b_i +c \delta x_i$,
\begin{equation}
      \tilde{x}_{i+1} = A^{-1}b_i + cA^{-1}\delta x_i,
\end{equation}
where $b_i$ is defined through the classical numerical solution $x_i$.  Making use of the triangular inequality one gets the following estimate for the fidelity
\begin{eqnarray}
\begin{gathered}
|x_{i+1}^\dagger \widetilde{x}_{i+1}|\geq \Bigl| \bigl|\underbrace{x_{i+1}^\dagger A^{-1}b_i}_{\sqrt{1-\epsilon_i}}\bigr|-c\bigl|x_1^\dagger A^{-1}\delta x_i\bigr|\Bigr| 
\geq \sqrt{1-\epsilon_i}-c\kappa(A)\bigl|x_{i+1}^\dagger\delta x_i\bigr|\approx
1-\frac{\epsilon_i}{2}\left(5+c\right).
\end{gathered}
\label{Error_fidelity_next_step}
\end{eqnarray}
Therefore, one gets an upper estimate for the error at the $i$th time step as $\epsilon_{i+1} \leq (5+c) \epsilon_i \leq (5+c)^i\epsilon_1$. Here $\epsilon_1$ is the error due to the finite depth of the variational ansatz. Thus as far as the number of time partitions $N_\tau\propto c$ increases the error of the variational algorithm scales as $\mathcal{O}\left((5+c)^{N_\tau-1}\right)$. Note, that the same picture is valid for the classical numerical solution: the arithmetic error scales as $\mathcal{O}\left((4+c)^{N_\tau-1}\right)$ with $c$, see Eq.(\ref{error_conditional_number}).

Let us fix the same accuracy of the final solution for quantum and classical schemes. In order to achieve this accuracy one searches for the optimal value of the grid parameter $c$ by minimizing  together the time derivative error $\propto \mathcal{O}(1/c)$ and the numerical scheme error $\propto \mathcal{O}\left((4+c)^{N_\tau-1}\right)$ or $\mathcal{O}\left((5+c)^{N_\tau-1}\right)$ in the classical or quantum case respectively.  One then checks whether or not  the polynomial regime of the ansatz depth $d \propto poly(n)$ is satisfied at this $c$ (see Figs. \ref{m_part2} and \ref{dependence_depth_n_qubits_fidelity99_ansatz_tree_approach}). Once the polynomial regime is observed the quantum scheme indeed outperforms the classical one.

\section{Conclusion}

The paper studies the implementation of three variational quantum algorithms for solving the heat equation presented in the finite difference form. This problem is reduced to the solution of the system of linear equations arising at each discrete step of the time evolution. 

In the  first approach (direct variational method) the expectation value of the Hamiltonian (\ref{var_ham}) is minimized on some class of probe functions. The Hamiltonian is constructed in a way that its ground state corresponds to the solution of the system of linear equations. We performed proof-of-principles quantum computation with the matrix of size $4\times 4$ using the real quantum processor of IBM Q project. The direct variational algorithm demonstrates a fundamental possibility of solving the system of linear equations (\ref{Ax=b}) on a quantum computer. However, the exponential number of Pauli products in the matrix decomposition does not allow one to achieve the quantum speedup (superiority over classical algorithms). In some cases it is possible to effectively sample over these products if we know the distribution of the decomposition coefficients, but this requires a separate study.

The second approach (Hadamard test approach) is based on the minimization of the expectation value of the same Hamiltonian, but the problem of the exponential number of Pauli products is eliminated by using the Hadamard test \cite{huang2021near}. A numerical simulation of the algorithm was performed with up to $n=8$ qubits using three different entanglers or ansatzs. The results show that it can be possible to achieve the quantum superiority, but the simulations with more qubits are required to definitively confirm this issue. It is also important to identify an effective entangler for the investigated problem. With this approach, three types of ansatzes were tested: the Hardware Efficient, Checkerboard and the Digital-Analog ansatz. The best results were obtained for the Checkerboard ansatz, as it gives a more uniform entanglement. In addition, by increasing the grid parameter $c$, see (\ref{c_definition}), one decreases the number of required layers in the ansatz. An exponential acceleration of up to eight qubits was demonstrated for this entangler. However, we argue that the considered number of qubits is not enough for an unambiguous conclusion about the advantage of the algorithm over the classical one.

The third type of approach (ansatz tree approach) minimizes the $l_2$ norm (\ref{Ansatz_tree_approach_loss_function}), rather than the expectation value of the Hamiltonian. The algorithm is based on the unitary decomposition of the matrix (\ref{A_definition}). For the heat equation it turns out to be advantageous to switch to the Fourier representation by using the quantum Fourier transform. In the Fourier representation, the matrix becomes diagonal with a sinusoidal spectrum (\ref{A_spectrum}). Then we used a technique that allows us to replace the spectrum of this matrix by a piecewise-quadratic function, which, at the level of the original discretized problem, corresponds to the elimination of high-frequency oscillations of the solution, justified from the physical point of view. This makes it possible to radically reduce the number of Pauli products in the matrix decomposition. The simulation of the algorithm with up to 11 qubits was performed and the complexity of the algorithm was estimated. The complexity is determined by the depth of the algorithm. The results show that the depth starting from certain value saturates on the number of qubits for certain values of the grid parameter $c$. This reveals the fundamental ability of the ansatz tree approach to demonstrate the quantum superiority for the heat equation.

Thus, the third approach can be considered as the most promising. The reason is that ansatz tree approach makes use of the explicit form of the matrix (\ref{A_definition}), unlike the other algorithms discussed, which use the universal entanglers.

\begin{acknowledgments}
We acknowledge use of the IBM Quantum Experience for this
work. The viewpoints expressed are those of the authors and
do not reflect the official policy or position of IBM or the
IBM Quantum Experience team.
\end{acknowledgments}

\bigskip
\bibliography{references}

\appendix

\section{Decomposition into Pauli products of matrices with a polynomial spectrum}

In this appendix we prove that a diagonal matrix with a polynomial spectrum can be decomposed into $\mathcal{O}\left(\left(\log N\right)^s\right)$ Pauli products, where $s$ is the highest degree of the polynomial and $N=2^n$ is the size of the matrix $A$.

Let us first consider the case when the diagonal matrix $D$ has a spectrum
\begin{eqnarray}
D=\sum^{N-1}_{m=0}m^s\ketbra{m}{m}.
\label{diagonal_matrix_spectrum_appendix}
\end{eqnarray}
We introduce a bit representation of the number $m$ as $\sum_k2^km_k$ and use the fact that
\begin{eqnarray}
\ketbra{m}{m}=\frac{1}{2^n}\prod^{n-1}_{k=0}\otimes\left(
I+e^{i\pi m_k}Z
\right).
\label{diagonal_matrix_through_I_and_Z_appendix}
\end{eqnarray}
Replacing the projector in (\ref{diagonal_matrix_spectrum_appendix}) with (\ref{diagonal_matrix_through_I_and_Z_appendix}) turns the expression (\ref{diagonal_matrix_spectrum_appendix}) into a decomposition of Pauli products
\begin{eqnarray}
D=\sum_{p=0}^{N-1}h_p\Pi_p,
\label{diagonal_matrix_with_polynomial_spectrum_Pauli_string_decomposition}
\end{eqnarray}
where $\Pi_p$ refers to the Pauli product consisting of $I$ and $Z$ gates. The number $p$ in its binary representation has $0$ and $1$ where $I$ and $Z$ are applied. Let the number $p$ have nonzero elements in the bit representation with numbers $k_1$, $k_2$,$\dots,k_l$
\begin{eqnarray}
p=2^{k_1}+2^{k_2}+\dots+2^{k_l}.
\label{p_bit_decomposition}
\end{eqnarray}
Thus, the decomposition coefficient $h_p$ has the form
\begin{eqnarray}
\begin{gathered}
h_p=\frac{1}{2^n}\sum_mm^s\exp\left(i\pi m_{k_1}+i\pi m_{k_2}+\dots+i\pi m_{k_l}\right)\\=\frac{1}{2^n}\sum_{\widetilde{m}} \sum_{m_{k_1},m_{k_2},\dots,m_{k_l}}\left(2^{k_1}m_{k_1}+2^{k_2}m_{k_2}+\dots+2^{k_l}m_{k_l}+\widetilde{m}\right)^s\left(-1\right)^{m_{k_1}+m_{k_2}+\dots+m_{k_l}},
\label{h_p_form_Pauli_sting_decomposition_appendix}
\end{gathered}
\end{eqnarray}
where $\widetilde{m}$ is the remaining part of the bit form of the number $m$ after separating $m_{k_1}$, $m_{k_2},\dots,m_{k_k}$. It follows from the expression (\ref{h_p_form_Pauli_sting_decomposition_appendix}) that if $l > s$, then $h_p=0$. This conclusion leads us to the following:

\textit{
For a diagonal matrix with spectrum $p(m) =\sum _{i=0}^s\alpha_im^i$, the Pauli product decomposition contains only terms with at most $s$ $Z$ operators.
}

Thus, for a polynomial of degree $s$, the number of nonzero terms of the decomposition into Pauli products $h_p$ is $\mathcal{O}\left(\left(\log N\right)^s\right)$.

\section{The fidelity of the ATA solution in the presence of the depolarizing noise}
\begin{figure}[t]
\begin{center}
\includegraphics[width=0.6\linewidth]{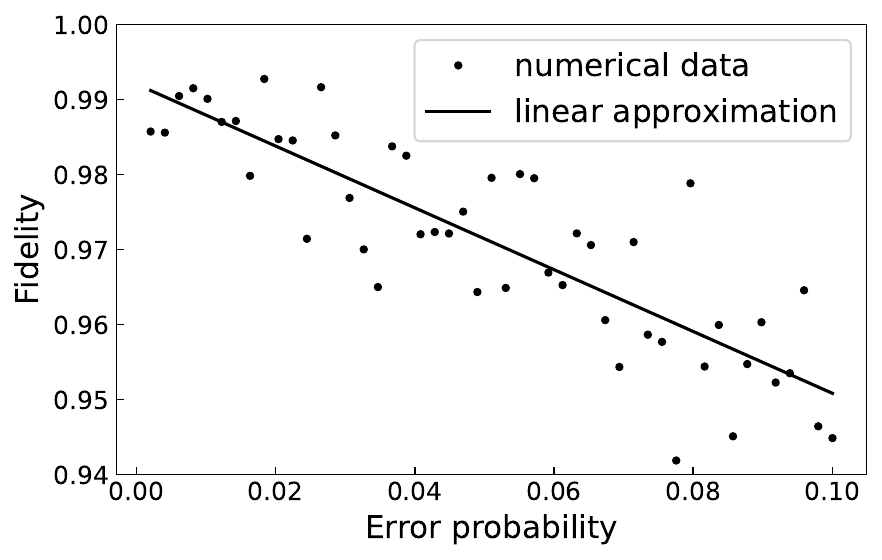}    
\end{center}
\caption{The dependence of the fidelity of the solution obtained by the ATA on the depolarizing noise parameter $p$ for a 4x4 matrix $A$. Each point is an average over 1000 initial conditions $\ket{b}$ and over 20 random values $c$ ranging from 0.1 to 2.0.}
\label{100_error_Ansatz_tree_2_qubits}
\end{figure}

This appendix examines the effect of depolarizing noise on the fidelity of the solution for the ansatz tree algorithm. The model of such a noise for one qubit is
\begin{eqnarray}
\mathcal{E}(\rho)=(1-p)\rho+p\frac{I}{2}.
\label{depolar_noise_Kraus}
\end{eqnarray}
In this noise model, the qubit state density matrix is replaced by the identical one $I/2$ with probability $p$. Figure \ref{100_error_Ansatz_tree_2_qubits} shows the effect of uncorrelated depolarizing noise on the implementation of an ansatz tree algorithm for two qubits. It can be seen from the graph that linear interpolation up to $p=1$ does not take a value close to the expected $1/2^n$, which corresponds to a random vector. To understand the reasons for this behavior, consider the effect of depolarizing noise on the ATA loss function
\begin{eqnarray}
L_R(x)=\sum_{j,k}\alpha_j^*\alpha_k\bra{j}A^2\ket{k}-2Re\{\alpha_j\sum_j\bra{j}A\ket{b}\}+1
\label{Loss_function_labeled_stated_asatz_tree_approach}
\end{eqnarray}
Convert the loss function to the form
\begin{eqnarray}
L_R(x)=\sum_{j}|\alpha_j|^2\bra{j}A^2\ket{j}-2Re\{\alpha_{0}\bra{0}A\ket{0}\}+1+\sum_{j\neq k}\alpha_j^*\alpha_k\bra{j}A^2\ket{k}-2Re\{\sum_{j\neq 0}\alpha_j\bra{j}A\ket{b}\}
\label{Loss_function_100_noise_labeled_stated_asatz_tree_approach}
\end{eqnarray}

Let us consider each contribution to the expression (\ref{Loss_function_100_noise_labeled_stated_asatz_tree_approach}). The last two sums of the expression (\ref{Loss_function_100_noise_labeled_stated_asatz_tree_approach}) are measured using the Hadamard test described in Section 4, the corresponding quantum circuit of such a measurement is shown in Fig.~ \ref{circ1}. In the presence of noise with $p=1$, each of the qubits will have a density matrix $I/2$ when measured. Consequently, due to the nature of the Hadamard test measurement, each term of these two sums has a mean value of zero. The first two terms of the expression (\ref{Loss_function_100_noise_labeled_stated_asatz_tree_approach}) are measured by the quantum circuit shown in Fig.~\ref{mean_measure_ansatz_tree}. These measurements contribute to the loss function
\begin{eqnarray}
L_R^{noise}(x)=\sum_{j}|\alpha_j|^2\bra{j}A^2\ket{j}-2Re\{\alpha_{0}\bra{0}A\ket{0}\}+1.
\label{truncated_Loss_function_100_noise_labeled_stated_asatz_tree_approach}
\end{eqnarray}
It can be seen from Eq. (\ref{truncated_Loss_function_100_noise_labeled_stated_asatz_tree_approach}) that the minimum corresponds to the condition $\alpha_i=\delta_{i0}$, then the solution of the algorithm is the state $\ket{0}=\ket{b}$.

\begin{figure}[t]
\begin{center}
\includegraphics[width=0.6\linewidth]{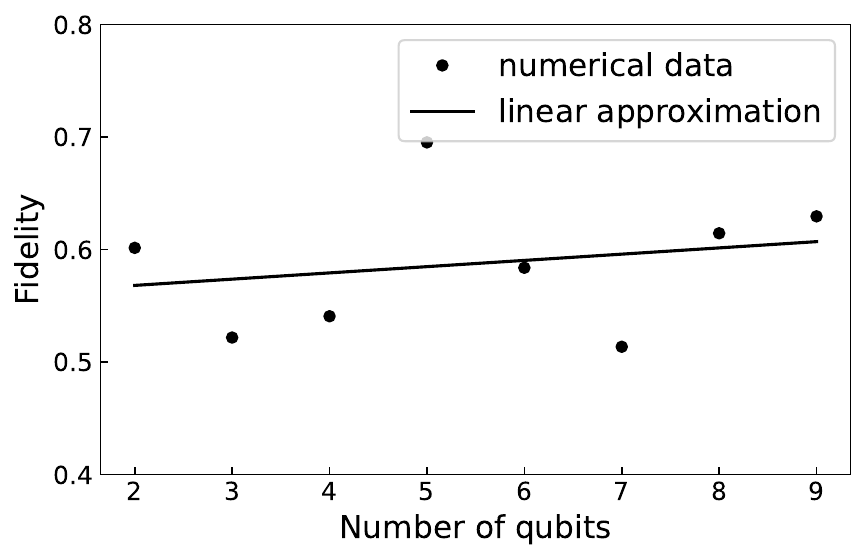}    
\end{center}
\caption{The dependence of $\left|\braket{b}{x}\right|^2$ (the fidelity of the ATA in the presence of noise at $p=1$) on the number of qubits (the size of the matrix $A$). Each point is an average over 1000 initial conditions $\ket{b}$ and 20 random values $c$ ranging from 0.1 to 2.}
\label{100_error_scaling}
\end{figure}

Thus, with noise for $p=1$, the algorithm converges not to $1/2^n$, but to $|\braket{b}{x}|^2$. Fig.~\ref{100_error_scaling} shows how the mean value of $|\braket{b}{x}|^2$ scales from the number of qubits (matrix size $A$). Each point on the graph corresponds to the average value of $|\braket{b}{x}|^2$ for 1000 different $\ket{b}$ and 20 random values of $c$ ranging from 0.1 to 2.

\end{document}